\documentclass[amsmath, amssymb, aps, prb, twocolumn, notitlepage]{revtex4-2}
\usepackage[utf8]{inputenc}
\usepackage{graphicx}               
\usepackage[caption=false]{subfig}  
\usepackage{dcolumn}                
\usepackage{bm}                     
\usepackage[dvipsnames]{xcolor}     
\usepackage{slashed}                
\usepackage{hyperref}               
\usepackage{enumitem}               
\usepackage{mathtools}              
\usepackage{floatrow}

\newcounter{myparagraphs}

\begin{document}
\title{Two-dimensional hydrodynamic electron flow through periodic and random potentials}

\author{Aaron Hui}
\thanks{These authors contributed equally to this work.}
\affiliation{Department of Physics, Ohio State University, Columbus, Ohio 43210, USA}
\author{Calvin Pozderac}
\thanks{These authors contributed equally to this work.}
\affiliation{Department of Physics, Ohio State University, Columbus, Ohio 43210, USA}
\author{Brian Skinner}
\affiliation{Department of Physics, Ohio State University, Columbus, Ohio 43210, USA}
\date{\today}

\begin{abstract}
We study the hydrodynamic flow of electrons through a smooth potential energy landscape in two dimensions, for which the electrical current is concentrated along thin channels that follow percolating equipotential contours. The width of these channels, and hence the electrical resistance, is determined by a competition between viscous and thermoelectric forces. For the case of periodic (moir\'{e}) potentials, we find that hydrodynamic flow provides a new route to linear-in-$T$ resistivity. We calculate the associated prefactors for potentials with $C_3$ and $C_4$ symmetry. On the other hand, for a random potential the resistivity has qualitatively different behavior because equipotential paths become increasingly tortuous as their width is reduced. This effect leads to a resistivity that grows with temperature as $T^{10/3}$. 
\end{abstract}

\maketitle

\emph{Introduction} -- Under conditions where electrons collide much more frequently with one another than with anything else, the current carried by an electron system flows like a fluid rather than satisfying the usual Ohm's law. This hydrodynamic electron regime was described by Gurzhi in the 1960s \cite{gurzhi_minimum_1963, gurzhi_hydrodynamic_1968}, and it has attracted significant attention during the last decade owing largely to its realization in graphene \cite{Lucas2018, Narozhny2022, deJong1995, Muller2009, Torre2015, Levitov2016, Bandurin2016, Crossno2016, Guo2017, Kumar2017, Bandurin2018, Braem2018, Hui2020, Bergdyugin2019, Gallagher2019, Sulpizio2019, Jenkins2020, Ku2020, Vool2021, Aharon-Steinberg2022, Moll2016, Bachmann2022, Gooth2018, Gusev2018, Levin2018, Gusev2020, Shavit2019, Stern2021, Kumar2022, valentinis2023, Gall2023, Gall2023a, Hui2021, Lucas2018b}. Recent experiments have demonstrated a variety of transport phenomena associated with hydrodynamic electrons, including negative non-local resistance \cite{Levitov2016, Bandurin2016, Bandurin2018, Braem2018, Levin2018, Samaddar2021, Hui2020}, Pouiselle-like flow profiles \cite{Sulpizio2019, Jenkins2020, Ku2020, Vool2021, Aharon-Steinberg2022, Gusev2020, Huang2023}, superballistic flow \cite{Guo2017, Kumar2017}, Wiedemann-Franz law violations \cite{Muller2008, Foster2009, Principi2015, Crossno2016, Lucas2016, Gooth2018, Lucas2018a, Zarenia2019, Zarenia2020, Robinson2021, Ahn2022, Li2022, Hui2023}, and bulk field expulsion \cite{Shavit2019, Stern2021, Kumar2022}. 

\begin{figure}
    \centering
    \includegraphics[width=.9\columnwidth]{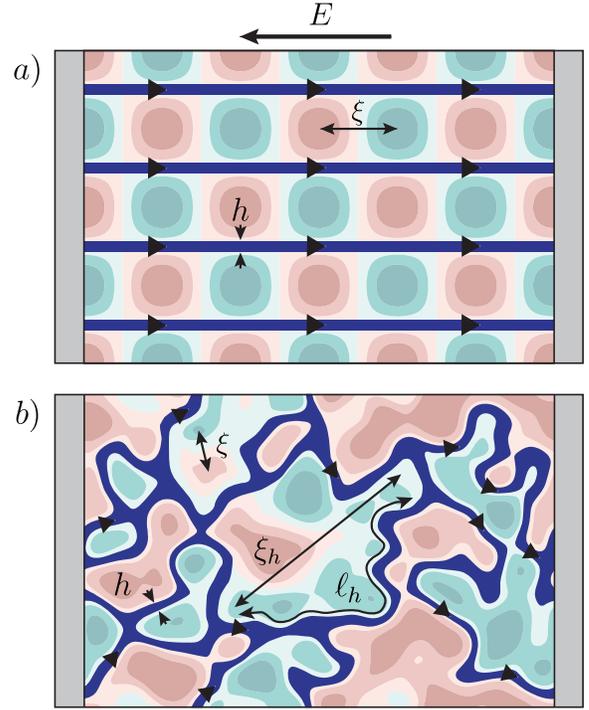}
    \caption{A device sketch with the imposed external potential $U$ (blue and red contour lines), where the shaded dark blue lines correspond to current flow under the applied electric field $E$. The thin current channels of width $h$ are concentrated about equipotential contours of $U$. a) The case of square periodic potential $U = U_0 \cos(2\pi x/\xi) \cos (2\pi y/\xi)$. b) The case of a random potential, for which the equipotential contours and the current channels meander and become tortuous in nature, controlled by the hull correlation length $\xi_h$ and the hull perimeter $\ell_h$.}
    \label{fig: BCs}
\end{figure}

Where disorder effects are included in descriptions of hydrodynamic electron flow, these effects are usually implemented via a finite momentum relaxation rate. Such a description is equivalent to imagining spatially uncorrelated, delta-function scatterers. On the other hand, in Ref.~\cite{Andreev2011} Andreev, Kivelson, and Spivak (AKS) considered hydrodynamic electron flow through a smooth random potential that varies on a length scale that is long compared to the electron-electron mean free path $\ell_{ee}$. AKS considered two contributions to the electrical resistance in this setting, arising from viscous shear stresses and thermoelectric fields. Using an ``energy minimization'' argument (properly, entropy maximization, as we explain below), they argued that when the electronic viscosity or thermal conductivity is low enough, the electric current in two dimensions is concentrated along narrow channels that follow equipotential contours, as sketched in Fig.~\ref{fig: BCs}. AKS derived a corresponding result for the resistivity (up to numeric prefactors). 

In this paper, we reconsider the problem of hydrodynamic flow through a smooth potential and provide two important updates to the AKS result. First, we consider the flow through a periodic (moir\'e) potential. We derive the corresponding resistivity, which follows the same form as the AKS result, and we give appropriate numeric prefactors for periodic potentials with $C_3$ and $C_4$ symmetry. We further show that, for electron systems obeying Fermi liquid theory, the result implies a linear-in-$T$ dependence of the resistance. 

These results may have a direct connection to recent transport experiments. Strong, slowly-varying periodic potentials now abound experimentally due to the explosion of interest in moir\'e systems \cite{Bistritzer2011, Cao2018b, Cao2020, Codecido2019, Yankowitz2019, Sharpe2019, Andrei2020, Balents2020, Jaoui2022, Wu2018, Wu2019b, Xian2019, Mak2022}. In both twisted bilayer graphene \cite{Polshyn2019} and TMD (transition metal dichalcogenide) systems \cite{Li2021, Ghiotto2021}, regimes of linear-in-$T$ resistivity have been experimentally discovered near strongly correlated phases. As both pedestrian explanations and exotic conjectures have been put forth for this temperature dependence \cite{Wu2019, Yudhistira2019, Cao2020b, DasSarma2020, DasSarma2022}, it is important that we understand all possible routes to linear-in-$T$ resistivity. 

Second, we turn our attention to the case of a spatially random potential. We show that the AKS result no longer applies because current-carrying channels become increasingly tortuous as their width decreases. Instead, the resistivity is governed by nontrivial critical exponents associated with two-dimensional (2D) percolation, leading to a superlinear $T^{10/3}$ dependence of the resistivity on temperature. We conclude with some brief remarks on how both results may be tested experimentally.

\emph{Mathematical Setup} -- The hydrodynamic equations that govern viscous electron flow are
\begin{align}
    &-\nabla P - en \mathbf{E} - \nabla U_\text{dis} - mn\nu \nabla \times \nabla \times \mathbf{v}
    \label{eq: momentum full}
    \nonumber
    \\
    &\phantom{-\nabla P} + mn\left[\frac{D-1}{D}2\nu + \tilde{\zeta}\right] \nabla \nabla \cdot \mathbf{v} = mn \mathbf{v}\cdot\nabla \mathbf{v}
    \\
    &\kappa \nabla^2 T + \frac{1}{2} mn\nu \left(\partial_i v_j + \partial_j v_i - \frac{2}{D} \delta_{ij}\partial_k v_k\right)^2 
    \label{eq: heat full}
    \nonumber
    \\
    &\phantom{\kappa \nabla^2 T + \frac{1}{2} mn\nu \Big(\partial_i v_j} + mn \tilde{\zeta} (\nabla \cdot \mathbf{v})^2 = mnT \mathbf{v}\cdot \nabla s
    \\
    &\nabla \cdot (n\mathbf{v}) = 0,
    \label{eq: density full}
\end{align}
where $m$ is the hydrodynamic mass, $-e$ is the electron charge, and $D = 2$ is the dimensionality. The hydrodynamic variables are the velocity $\mathbf{v}$, the pressure $P$, the particle density $n$. We treat the electric field $\mathbf{E}$ as a weak, externally applied field. Eq.~\eqref{eq: momentum full} is the Navier-Stokes (momentum) equation, with kinematic shear viscosity $\nu$ and kinematic bulk viscosity $\tilde{\zeta}$, as well as the externally imposed disorder potential $U_\text{dis}$. Eq.~\eqref{eq: heat full} is the heat (energy) equation, with thermal conductivity $\kappa$ and entropy per unit mass $s$. Finally, Eq.~\eqref{eq: density full} is the density continuity equation. To complete the set of equations, we need constitutive relations between our hydrodynamic variables. Since $(s,T)$ and $(n,P)$ are thermodynamically conjugate variables, we choose one from each set to be our independent variables. In particular, we choose variables $n$ and $T$ so that
\begin{align}
    \nabla P(n,T) =& \frac{\partial P}{\partial n} \nabla n - mn^2 \frac{\partial s}{\partial n} \nabla T
    \label{eq: P constitutive relation}
    \\
    \nabla s(n,T) =& \frac{\partial s}{\partial n} \nabla n + \frac{\partial s}{\partial T} \nabla T
\end{align}
where $n_s \equiv mns$ is the entropy density and we used the thermodynamic relation $(\partial P/\partial T) = - mn^2 (\partial s/\partial n)$. For simplicity, we assume that $\partial P/\partial n >0$ and $\partial s /\partial n<0$ are constants. Finally, we consider a rectangular domain $[0,L_x]\times [0,L_y]$ as shown in Fig.~\ref{fig: BCs}. For boundary conditions (BCs), we fix $T=\overline{T}$ and take periodic BCs for $n$ on the $x$-boundaries. Furthermore, we take for simplicity periodic BCs on the $y$-boundaries \footnote{For strong disorder where the currents are isolated to thin channels (e.g. in Fig.~\ref{fig: BCs}), we expect the choice of $y$-BC to only be relevant near the boundary. This is because we expect the localized current channels to well approximated by channels obeying no-slip conditions (see Fig.~\ref{fig: variational} and the surrounding discussion).}.  

We are interested in the linear-response theory of the above equations without assuming that $\nabla U_\text{dis}$ is weak. Therefore, we look to organize our solution in a formal perturbative scheme $\mathbf{v} = \mathbf{v}^{(0)} + \mathbf{v}^{(1)} + \ldots$, and similarly for the other hydrodynamic variables. We will determine the explicit non-dimensionalized perturbative parameter \textit{ex post facto}. Of course, physically the expansion is controlled by the perturbatively weak electric field $\mathbf{E}$. At leading (zeroth) order, we consider the equilibrium situation where we expect $\mathbf{v}^{(0)} = 0$ and $T^{(0)} = \overline{T}$. Therefore, the only non-trivial equation at zeroth order is 
\begin{align}
    -\nabla P^{(0)} - \nabla U_\text{dis} = 0
\end{align}
where we have kept $U_\text{dis}$ since it is not perturbatively small. From the constitutive relations, this implies that $\nabla n^{(0)} \propto \nabla U_\text{dis} \propto \nabla s^{(0)}$. Thus, the density and entropy per mass profiles are inherited from the disorder potential at leading order.

We now consider the first-order hydrodynamic equations, driven by a perturbatively weak field $\mathbf{E}$. These are given by the equations
\begin{align}
    &-\nabla P^{(1)} - en^{(0)} \mathbf{E} - mn^{(0)}\nu \nabla \times \nabla \times \mathbf{v}^{(1)}
    \nonumber 
    \\
    & \phantom{-\nabla P^{(0)}} + mn^{(0)} \left[\frac{D-1}{D}2\nu + \tilde{\zeta}\right]\nabla\nabla\cdot\mathbf{v}^{(1)} = 0
    \label{eq: momentum}
    \\
    &\kappa \nabla^2 T^{(1)} = m n^{(0)} \overline{T} \mathbf{v}^{(1)} \cdot \nabla s^{(0)}
    \label{eq: heat}
    \\
    &\nabla \cdot (n^{(0)} \mathbf{v}^{(1)}) = 0,
    \label{eq: continutity}
\end{align}
where we treat $\mathbf{E}$ as a first-order perturbation. 
Eqs.~(\ref{eq: momentum}) -- (\ref{eq: continutity}) are equivalent to those in Ref.~\onlinecite{Andreev2011}, with the perturbation theory considerations manifestly written. It is crucial that one utilizes the temperature-dependence in Eq.~\eqref{eq: P constitutive relation}; otherwise, Eq.~\eqref{eq: momentum} decouples from Eq.~\eqref{eq: heat}. This dependence provides a ``thermoelectric'' contribution to Eq.~\eqref{eq: momentum}, which is the key term in restricting current to flow along narrow channels \footnote{In Ref.~\onlinecite{Andreev2011}, they argue that flow must be concentrated along equipotential lines in the $\kappa \rightarrow 0$ limit because the LHS of Eq.~\eqref{eq: heat} vanishes. However, this argument requires care because sending $\kappa \rightarrow 0$ is a singular operation; because $\kappa$ acts on the highest derivative, $\kappa \rightarrow 0$ is \emph{not} generally equivalent to $\kappa = 0$. Alternatively, $\kappa \rightarrow 0$ does not necessarily mean the LHS of Eq.~\eqref{eq: heat} vanishes since one would also need to prevent $\nabla^2 T^{(1)}$ from growing arbitrarily large. Without the temperature-dependent contribution of Eq.~\eqref{eq: P constitutive relation}, $\nabla^2 T^{(1)}$ will diverge everywhere in the $\kappa \rightarrow 0$ limit to satisfy Eq.~\eqref{eq: heat}. The proper inclusion of the ``thermoelectric term'' provides a feedback loop that prevents this divergence.}.

A convenient way to obtain the two-terminal resistance $R$ is to compute the total entropy generation. The relation between these two quantities is subtle, and proceeds as follows. One can show that the entropy production of a hydrodynamic system is given by \cite{landauv6}
\begin{align}
    &\int dV \frac{d n_s}{dt} = -\oint \left(n_s \mathbf{v} + \frac{\kappa \nabla T}{T}\right)\cdot d\mathbf{A} + \int dV \frac{q}{T}
    \label{eq: entropy generation}
\end{align}
with    
\begin{align}
    &q \equiv \frac{1}{T}\kappa (\nabla T)^2 + \frac{1}{2} mn\nu \left(\partial_i v_j + \partial_j v_i -\frac{2}{D}\delta_{ik} \partial_l v_l\right)^2 
    \nonumber
    \\
    &\phantom{q = \frac{1}{T}\kappa (\nabla T)^2} + mn \tilde{\zeta} (\nabla\cdot \mathbf{v})^2.
    \label{eq: q}
\end{align}
Note that $q/T$ is positive semi-definite and can therefore be interpreted as the bulk entropy production. In steady-state the LHS of Eq.~(\ref{eq: entropy generation}) vanishes, and thus all the bulk-generated entropy flows out through the contacts held at $\overline{T}$. On physical grounds, we assume that this entropy outflow is gained as heat by the environment through the contacts at temperature $\overline{T}$. Thus, by equating the dissipated $I^2 R$ power to the environmental heating, we have
\begin{align}
    I^2 R &= \overline{T} \int dV \frac{q}{T}
    \label{eq: I2R}
\end{align}
When the variations of $T$ are small such that $T \approx \overline{T}$, we have the simpler relation $I^2 R = \int dV q$ as written by Ref.~\cite{Andreev2011}. Only in this limit of $\delta T \ll \bar{T}$ can one interpret Eq.~\eqref{eq: I2R} as energy conservation with $q$ as the ``local power dissipation'' \footnote{A heat current cannot dissipate energy; by definition it is a conserved current of energy. Consider, for instance, an insulated metal plate with a non-uniform temperature distribution. The total energy of the plate is always conserved, yet heat currents flow. Instead, the plate maximizes its entropy as it relaxes towards equilibrium.}. Throughout this paper, we make the assumption $\delta T \simeq T^{(1)} \ll \overline{T}$ (placing a bound on the true perturbative parameter $\mathbf{E}$) and therefore use the simpler relation.

\emph{Periodic Potential} -- Using Eqs.~(\ref{eq: q}) and (\ref{eq: I2R}), we calculate the resistance for different cases of the disorder potential. Let us first consider the case of a square periodic potential ${U_\text{dis, sq} = U_0 \cos(2\pi x/\xi) \cos (2\pi y/\xi)}$ with periodicity $\xi$ (see Fig.~\ref{fig: BCs}a); this case was sketched by Ref.~\onlinecite{Andreev2011}. As we argued above, the zeroth order density $n^{(0)}$ and entropy density $s^{(0)}$ also fluctuate around their mean values with the same spatial periodicity. In the strong disorder limit, we make the ansatz that the flow is isolated to thin horizontal channels of width $h$ and length $\ell = L_x$, centered around the equipotential lines of $s^{(0)} = \overline{s}$ (see Fig.~\ref{fig: BCs}a). Each of the $N = L_y/\xi$ such channels carries an equal amount of current $I/N$, where $I$ is the total current. We further assume that the flow is incompressible, i.e.\ that $\nabla \cdot \mathbf{v} = 0$. This incompressibility assumption is justified if $(n^{(0)} - \overline{n})\ll \overline{n}$ within the channel \footnote{As a technical note, we also must assume that flow velocity $v \ll c$, where $c$ is the speed of sound. This ensures that $n^{(0)} \gg n^{(1)}$ \cite{landauv6}.}; we show below that this assumption is valid for $h/\xi \ll 1$. Finally, we assume that the temperature fluctuations outside of the channel are negligible, since the dominant heating is isolated to within the thin channels.

Assuming that the flow chooses an optimum channel width $h$ to minimize the total dissipated power, we estimate the power dissipation. Implicit in this assumption is that the heat current influences flow, e.g.\ through a thermoelectric term. In the incompressible limit, the leading order contribution to dissipation is
\begin{align}
    I^2 R = N \int_\text{ch} dV \frac{1}{T_0} \kappa \left(\nabla T^{(1)}\right)^2 + \frac{m \overline{n}\nu}{2} \left(\partial_i v_j^{(1)}
    + \partial_j v_i^{(1)}\right)^2,
    \label{eq: power}
\end{align}
where the integral is over a single channel and we can approximate $n^{(0)} \sim \overline{n}$. By a scaling estimate similar to the one in Ref.~\cite{Andreev2011}, we find
\begin{align}
    I^2 R \sim \frac{I^2}{N e^2} \frac{\ell}{\xi} \left[\frac{\overline{T}}{\kappa} (m \overline{\delta s})^2 \left(\frac{h}{\xi}\right)^3 + \frac{m \nu}{\overline{n} \xi^2} \left(\frac{\xi}{h}\right)^3\right]
    \label{eq: power_scalings}
\end{align}
where $\overline{\delta{s}}$ is the characteristic amplitude of the entropy fluctuations and we have used Eq.~\eqref{eq: heat} and the approximations $h\ll \xi$, $v_y \ll v_x$, $\partial_x \sim 1/\xi$, and ${\partial_y \sim 1/h}$. From Eq.~\eqref{eq: power_scalings} one can see that there are two resistance contributions which compete in determining the channel width $h$. The first term, corresponding to dissipation from thermoelectrically-driven heat currents, favors narrow channels. The second term, corresponding to dissipation from viscous shearing, favors wide channels. Minimizing the dissipated power against $h$, we find the channel width to scale as
\begin{align}
    \frac{h}{\xi} \sim& \, \alpha^{-1/6},
    \label{eq: width}
\end{align}
where
\begin{align}
    \alpha \equiv& \frac{\overline{T} \overline{\delta n_s}^2 \xi^2}{\kappa \eta}
\end{align}
is the ratio of ``thermal'' to viscous dissipation [see Eq.~\eqref{eq: power_scalings}], $\overline{\delta n_s} = m \overline{n} \overline{\delta s}$ is the characteristic strength of entropy density fluctuations and $\eta = m \overline{n} \nu$ is the dynamic viscosity. Therefore, we find perturbative control when $\alpha \gg 1$ (and, correspondingly, channels are narrow). Furthermore, we need to ensure that the thermoelectric term in Eq.~\eqref{eq: P constitutive relation} is sufficiently large to ensure that channels actually form. A perturbative solution around  $\partial s/\partial n = 0$ does not form channels; since Eq.~\eqref{eq: heat} decouples in this limit, the solution has non-zero velocity everywhere with velocity variations set by $\xi$ from the continuity equation [Eq.~\eqref{eq: continutity}]. Via a scaling estimate, this perturbative ansatz fails when $ {(\overline{n}/\overline{\delta s})|\partial s/\partial n| \alpha \gg 1}$. Finally, our incompressibility assumption is valid if $(\overline{\delta n}/\overline{n}) \alpha^{-1/6} \ll 1$ for $\overline{\delta{n}}$ the characteristic strength of density fluctuations. Thus, all our assumptions are controlled by $\alpha \gg 1$ up to dimensionless factors. We note that the explicit bounds on the validity of the thin channel ansatz was not previously treated by AKS; in particular, a non-trivial thermoelectric term (i.e. $\partial s/\partial n$) is necessary.

Plugging Eq.~\eqref{eq: width} into Eq.~\eqref{eq: power_scalings}, we find the resistivity to be \footnote{Throughout this paper, we define the (effective) resistivity $\rho = R L_y/L_x$; it is important to keep in mind that by resistivity, we do \emph{not} mean that a Ohm's law relation $\rho J = E$ holds.}
\begin{align}
    \rho \sim \frac{2}{e^2}\sqrt{\frac{\overline{T} \eta (m\overline{\delta s})^2}{\kappa\overline{n}^2 \xi^2}}
    \label{eq: square periodic scaling resistivity}
\end{align}
This equation recovers the results of Ref.~\onlinecite{Andreev2011}. Below we numerically verify these results and determine the proportionality coefficient [see Eq.~\eqref{eq: numerical resistivity}]. For a Fermi liquid, Eq.~\eqref{eq: square periodic scaling resistivity} implies a particular temperature dependence of the resistance. Specifically, a Fermi liquid has viscosity $\eta \sim T^{-2}$, thermal conductivity $\kappa \sim T^{-1}$, and entropy density $\overline{\delta n_s} \sim T$ \cite{Andreev2011}. These substitutions give $h \propto 1/T$ and we find a linear $\rho \propto T$ scaling, as mentioned above.

\begin{figure}[tb!]
\begin{center}
\includegraphics[width=1\columnwidth]{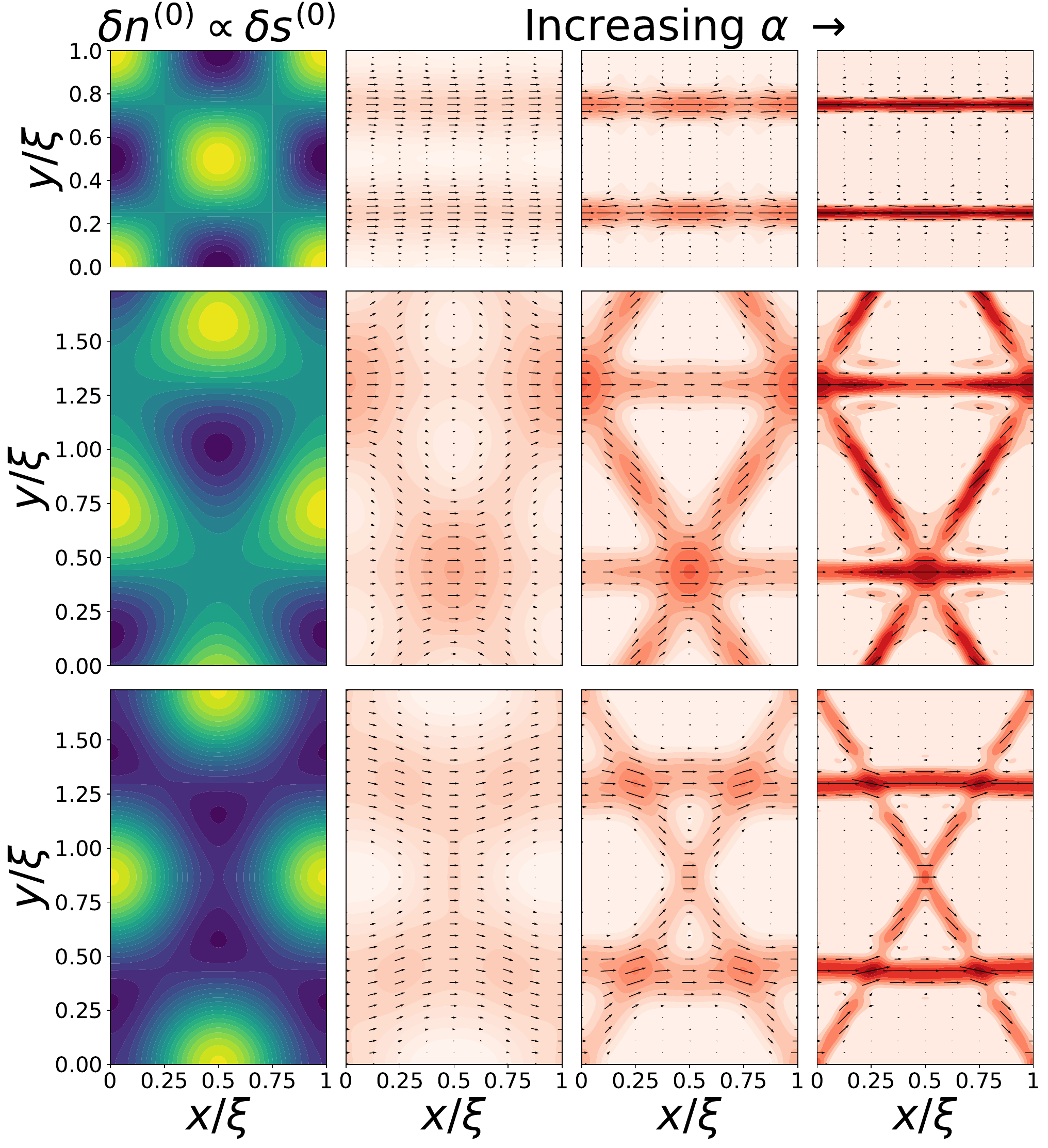}
\end{center}
    \caption{Numerical solutions for the current density $\mathbf{j}$ in various periodic potentials. Each row corresponds to a particular type of potential, illustrated in the first column (top to bottom: square potential, triangular potential with $\psi = \pi/6$, triangular potential with $\psi = 0$). Columns 2-4 show the simulated current density (arrows show direction and color darkness shows magnitude) for each potential. Columns 2-4 correspond to $\alpha \sim 10^3, 10^5,$ and $10^7$, respectively. As $\alpha$ increases, the current-carrying channels become increasingly narrow. 
    }
    \label{fig: simulations}
\end{figure}

\emph{Numerical Simulation} -- For the case of periodic potentials, we can provide direct numerical solutions of the hydrodynamic equations to verify our scaling results. Specifically, we solve Eqs.~(\ref{eq: momentum}) -- (\ref{eq: continutity}) with the above BCs using the spectral PDE solver Dedalus \cite{2020PhRvR...2b3068B}. We emphasize that for these simulations we make no assumptions about $n^{(0)}$ and in particular do not assume incompressibility. For simplicity, we assume the bulk viscosity $\tilde{\zeta} = 0$ in our simulations; numerically tuning this parameter has little effect on the qualitative flow profile. This irrelevance of $\tilde{\zeta}$ is as expected, since we expect flow to be approximately incompressible when thin channels form. In addition to the square potential, we consider a class of triangular potentials that describe the moir\'e pattern arising from mismatched hexagonal lattices (as in graphene or transition metal dichalcogenides) \cite{Wu2018}. Such potentials have one free parameter, $\psi$, that describes the phase difference between the moir\'e reciprocal lattice vectors (see the Supplemental for details). The results of these numerical simulations are shown in Fig. \ref{fig: simulations} for a range of values of $\alpha$. We observe the formation of current-carrying channels along the equipotential contours that span the system. Furthermore, the channels become increasingly narrow as $\alpha$ is increased, as predicted. 

\begin{figure}
\begin{center}
\includegraphics[width=0.9\columnwidth]{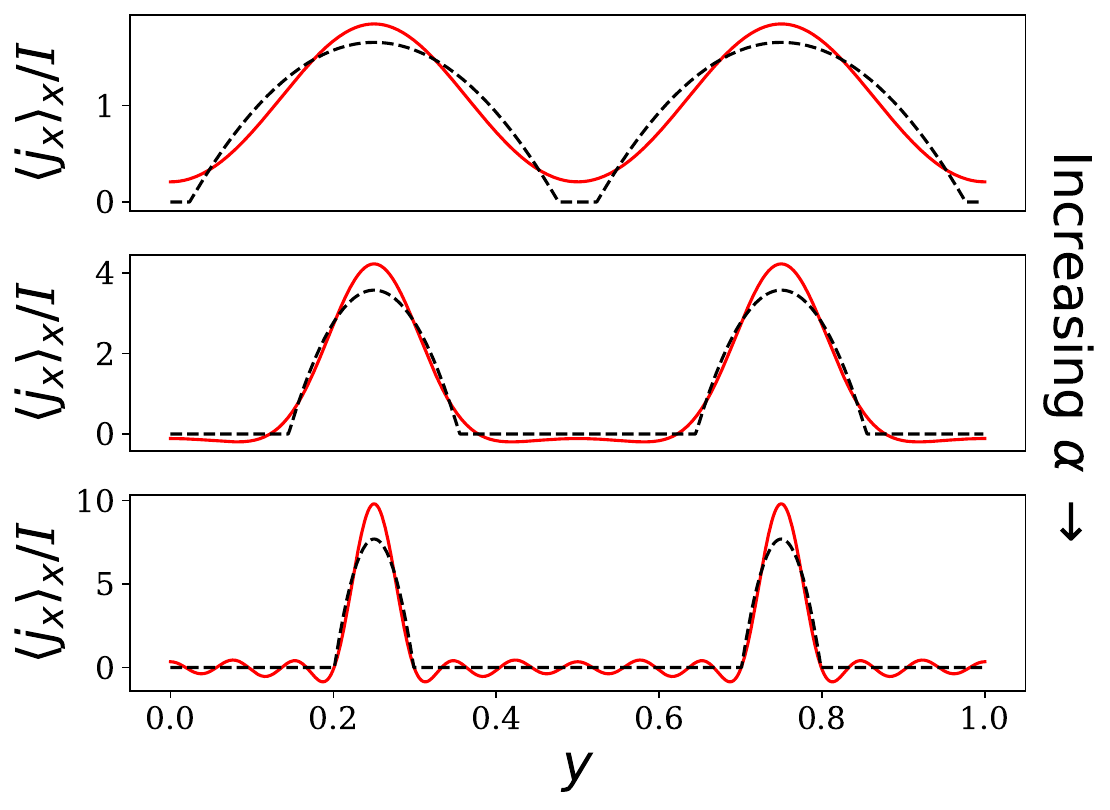}
\end{center}
    \caption{Numerical results for the averaged current density profile $\langle j_x \rangle_x \equiv \int dx j_x/L_x$ in a square-periodic potential, normalized by total current $I$. The three values of $\alpha$ are the same as those in Fig.~\ref{fig: simulations}. The solid lines correspond to the results of direct numerical simulation, while dashed lines correspond to our approximate variational solution. }
    \label{fig: variational}
\end{figure}

In order to provide a quantitative calculation of the resistance, we adopt a variational approach that assumes a parabolic flow profile within each channel. Specifically, we assume a current density $j_x(x,y) = 6(I/N)[(h/2)^2-y^2]/h^3$ within each channel (with $y=0$ corresponding to the center of a given channel) and zero elsewhere. The width $h$ of the channel is treated as a variational parameter; see Fig.~\ref{fig: variational} for a comparison between our ansatz for $j_x(y)$ and exact numerical solutions. This ansatz for $\mathbf{j} = e \overline{n} \mathbf{v}$ yields a temperature $T^{(1)}$ via Eq. (8). Consequently, we arrive at analytic expressions for both of the power dissipation terms in Eq.~\eqref{eq: power_scalings}, in the limit of $h\ll \xi$, with exact numerical prefactors for square and triangular potential profiles: 
\begin{align}
    \label{eq: square_scaling_th}
    Q_{\text{th}}  &= C_\text{th}\frac{I^2 L_x \overline{T}}{e^2 L_y \kappa} (m \overline{\delta s})^2 \left(\frac{h}{\xi}\right)^3 \\
    Q_{\text{vis}} &= C_\text{vis}  \frac{ I^2 L_x m \nu}{e^2 L_y \overline{n} \xi^2} \left(\frac{\xi}{h}\right)^3 
    \label{eq: square_scaling_vis}
\end{align}
\begin{alignat}{3}
    C_\text{th, sq} &= \frac{\pi^4}{35}   &&,\:\:  C_\text{th, tri} & &=\frac{4\pi^4}{630}\\ 
    C_\text{vis, sq} &= 24   &&, \:\:C_\text{vis, tri} & &=\frac{24}{\left(1-\frac{\overline{\delta n}}{\sqrt{6}\,n}\text{cos}(3\psi)\right)^2}
    \label{eq: coeffs}
\end{alignat}
where $Q_\text{th}$ and $Q_\text{vis}$ correspond to the thermal (first) and viscous (second) terms in Eq.~\eqref{eq: power_scalings}.

As before, we look for a channel width $h$ such that $Q_{\text{th}}+Q_{\text{vis}}$ is minimized.
We plot the variational result for the current density in Fig.~\ref{fig: variational} along with the corresponding result from direct numerical simulation, which shows close agreement. Finally, we compute the resistivity by evaluating the total power with the variationally-determined channel width $h$. This procedure gives
\begin{align}
    \rho  = \frac{C}{e^2}\frac{\overline{\delta n_s}}{ \overline{n}^2\xi}\sqrt{\frac{\overline{T} \eta}{\kappa}}.
    \label{eq: numerical resistivity}
\end{align}
This result validates the scaling result of Eq.~\eqref{eq: square periodic scaling resistivity} up to the numerical prefactor $C$, which for square and triangular potentials are given by
\begin{align}
    C &= 4 \pi^2 \sqrt{\frac{6}{35}} & \textrm{(square)}  \\ 
    C &= \frac{8 \pi^2}{\sqrt{105}\left(1-\frac{\overline{\delta n}}{\sqrt{6}\,n}\text{cos}(3\psi)\right)} & \textrm{(triangular)}
\end{align}
(see the Supplemental for details).

\emph{Random Potential} -- We now turn our attention to the case of a smooth random potential with a correlation length $\xi$ \footnote{One possible method for constructing such a potential is as follows. We consider a disorder function of the form $U_{\text{dis}} (\mathbf{x}) \propto \sum_{\mathbf{k}\neq \mathbf{0}} A_{\mathbf{k}} \text{exp} \left(-k^2\xi^2/2\right) \text{cos}\left(\mathbf{k} \cdot \mathbf{x} +2\pi \delta_{\mathbf{k}} \right)$ where $A_{\mathbf{k}}$ and $\delta_{\mathbf{k}}$ are sampled randomly from $[0,1]$ with $A_\mathbf{k}$ normalized such that $\langle U_{\text{dis}} \rangle = U_0^2 $ and $\mathbf{k} = (\pi n_x/L_x, \pi n_y /L_y)$ with $n_x,n_y\in \mathbb{Z}$.}. Such random potentials arise, for example, from charged impurities in the substrate or an adjacent delta-doping layer, for which that the typical wave vector of the disorder potential is much smaller than the electron wave vector (see, e.g., Ref.~\cite{sammon2018mobility} and references therein). This consideration is distinct from a model of point defect scatters studied in, e.g., Ref.~\cite{guo2017stokes}. 

The key conceptual novelty of a random potential is that equipotential lines are very tortuous \cite{Isichenko1992}, and therefore so are the current-carrying channels (see Fig.~\ref{fig: BCs}b). In particular, the number of parallel current-carrying channels $N$ and the contour length $\ell$ of each channel now depend on percolation exponents. In 2D, the hull correlation length exponent is $\nu_h = \nu = 4/3$ and the hull perimeter exponent is $d_h = 7/4$ \cite{Isichenko1992}. Taking $\xi$ to be the disorder correlation length and $\xi_h$, $\ell_h$ to be the hull correlation length and hull perimeter, respectively, (see Fig.~\ref{fig: BCs}b) we have
\begin{align}
    \frac{\xi_h}{\xi} &\sim \left(\frac{\xi}{h}\right)^{4/3}
    \\
    \frac{\ell_h}{\xi} &\sim \left(\frac{\xi_h}{\xi}\right)^{7/4} \sim \left(\frac{\xi}{h}\right)^{7/3}
\end{align}
One can think that current-carrying channels form a random network, with $\xi_h$ being the typical spacing between neighboring nodes in the network and $\ell_h$ being the length of the tortuous links between nodes.

With these results, we can again minimize the dissipated power [Eq.~\eqref{eq: power}]; the only difference from the periodic case is that the number of channels $N \sim L_y/\xi_h$ and the channel length $\ell \sim (L_x/\xi_h) \ell_h$ have nontrivial dependencies on the channel width $h$. With these new estimates, we find
\begin{align}
    \frac{h}{\xi}\sim \left(\frac{\overline{T} \overline{\delta n_s}^2 \xi^2}{8\kappa \eta} \right)^{-1/6} \sim \alpha^{-1/6}.
\end{align}
Surprisingly, the channel width $h$ has the same scaling behavior as in the periodic case. However, the scaling behavior of the resistance is different, namely
\begin{align}
    \rho \sim \frac{2}{e^2}\sqrt{\frac{\overline{T} \eta (m\overline{\delta s})^2}{\overline{n}^2 \xi^2}} \left(\alpha^{1/6}\right)^{7/3}.
\end{align}
Thus we obtain a similar result as the periodic case [Eq.~\eqref{eq: square periodic scaling resistivity}] since $N$ and $\ell$ only provide an overall scaling factor of $(\xi/h)^{7/3}$ to the total power. Using the Fermi liquid scaling relations as before, we find $\rho \propto T^{10/3}$.

\emph{Conclusion} -- In this Letter, we have analyzed the resistance of hydrodynamic flow through both periodic and random smooth potentials. We find a novel mechanism for linear-in-$T$ resistance associated with hydrodynamic flow through a periodic potential, which we confirm by numeric simulations and variational calculations that allow us to precisely determine the relevant prefactors for square-periodic and triangular-periodic potentials. If systems can be made sufficiently clean, it may be possible to engineer moir\'e potentials to see such a linear-in-$T$ resistance, similar to what has been seen near strongly correlated phases of moir\'e systems \cite{Polshyn2019, Li2021, Cao2020b, Chu2022, Liu2020, Cao2020, Shen2020, He2021}. For generic random potentials, on the other hand, the tortuous nature of the current paths leads to a resistance temperature scaling of $T^{10/3}$. Such behavior may arise in a clean, hydrodynamic 2D electron system adjacent to a delta doping layer or a substrate with dilute charged impurities.

Throughout this paper, we have assumed rotational invariance and Galilean invariance of the fluid (excluding the smooth potential). We expect these assumptions to apply for electron systems with nearly-circular Fermi surfaces and at sufficiently low temperatures that only one band of carriers is relevant for conduction. In graphene and graphene multilayers, these assumptions usually translate to a Fermi energy $E_F$ that is low compared to the band width (for which the Fermi surface becomes nearly circular) and a temperature that is low compared to $E_F/k_B$. For specific moir\'e systems of interest, one may need to weaken these assumptions to include effects such as a non-circular Fermi surface \cite{Cook2019, Bachmann2022} and intrinsic (or incoherent) conductivity \cite{Foster2009, Muller2008, Hartnoll2015}. The loss of rotational symmetry promotes viscosity from a scalar to a tensor, and the inclusion of intrinsic conductivity will lead to an additional dissipation mechanism to compete with the ``thermal'' and viscous dissipation terms. We leave the treatment of these additional effects to future work.

\textbf{Acknowledgements} -- We thank Alex Levchenko and J.~C.~W.~Song for helpful discussions. C.~P.\ was supported by the Center for Emergent Materials, an NSF-funded MRSEC, under Grant No.\ DMR-2011876. B.~S.\ was partly supported by NSF Grant No.\ DMR-2045742.

\appendix

\section{Derivation of the resistivity numerical coefficients}

Here, we more carefully describe the periodic potentials we consider along with the derivation of the numerical coefficients. The exact forms of disorder, expressed through $n^{(0)}$ and $s^{(0)}$ are given by
\begin{alignat}{2}
    U_{\text{sq}} &= \cos &&\biggl(\frac{2\pi x}{\xi}\biggr) \sin \left(\frac{2\pi y}{\xi}\right)\\
    U_{\text{tri}} &= \frac{1}{\sqrt{6}}&&\biggl(2 \left(\cos \left(\frac{2 \pi  x}{\xi}\right) \sin \left(\frac{2 \pi  y}{\sqrt{3} \xi}\right)\right)\\
    & &&-\cos \left(\frac{4\pi y}{\sqrt{3} \xi}+3 \psi\right)\biggr).
\end{alignat}
This second equation defines the phase constant $\psi$ mentioned in the main text.
The fluctuations are normalized such that $\sqrt{\langle \delta U)^2\rangle} = 1/2$. The subsequent density and entropy per unit mass are $n^{(0)} = \overline{n^{(0)}} + \delta n U$ and $s^{(0)} = \overline{s^{(0)}} + \delta s U$, respectively.

In order to estimate the numerical coefficients of the resistivity for these potentials, we first estimate the coefficients of the viscous and thermoelectric power dissipation. We take an idealized approximation that the current flows in parabolic channels of width $h$ around the the spanning equipoitential contours. As described in the main text, we assume a current density 
\begin{equation}
j_x(x,y) = 6\frac{I}{N} \frac{(h/2)^2-y^2}{h^3} 
\end{equation}
within each channel (with $y=0$ corresponding to the center of a given channel) and zero elsewhere (see Fig.~\ref{fig: variational}).
This assumption along with a specific disorder potential immediately allows us to calculate the viscous dissipation and corresponding approximation of $T^{(1)}$ through Eq.~\eqref{eq: heat} to obtain the thermoelectric dissipation. Note that the continuity equation [Eq.~\eqref{eq: continutity}] is satisfied by this assumption while the Navier-Stokes equation [Eq.~\eqref{eq: momentum}] simply defines $\nabla P$ and thus can be disregarded for our purposes. 

For the viscous dissipation, the dominant contribution is given by:
\begin{align}
    Q_\text{vis} &= N \int_\text{ch} dV \frac{m \overline{n}\nu}{2} \left(\partial_i v_j^{(1)}
    + \partial_j v_i^{(1)}\right)^2\\
    &= 2N m \overline{n}\nu \int_\text{ch} dV \left(\partial_x \left(\frac{j_x}{e n^{(0)}}\right)
    + \partial_y \left(\frac{j_x}{e n^{(0)}}\right)\right)^2\\
    &\simeq \frac{2 m \overline{n}\nu L_y}{\xi} \int_{-h/2}^{h/2} dy \int_{0}^{L_x} dx\left(\partial_y \left(\frac{j_x}{e n^{(0)}}\right)\right)^2.
\end{align}
Taylor expanding $n^{(0)}(x,y)$ around $y=0$ allows us to integrate to obtain Eq.~(\ref{eq: square_scaling_vis}).

We now turn to the case of the thermoelectric dissipation. Here, we are left to solve
\begin{equation}
    \kappa \nabla^2 T^{(1)} = m\overline{T}(\mathbf{j}/e)\cdot\nabla s^{(0)} = m\overline{T}(j_x/e) \partial_x s^{(0)}
\end{equation}
in the channel and $\nabla^2 T^{(1)} = 0$ outside the channel with forced continuity and differentiablity of $T^{(1)}$ at the channel edges. This approximation for $T^{(1)}$, again expanded around $y = 0$, is then used to evaluate the dominant thermoelectric dissipative term:
\begin{align}
    Q_\text{th} &= N \int_\text{ch} dV \frac{1}{T_0} \kappa \left(\nabla T^{(1)}\right)^2\\
    & =  \frac{N\kappa}{T_0} \int_{-h/2}^{h/2} dy\int_{0}^{L_x} dx \left(\nabla T^{(1)}\right)^2\\
    & \approx  \frac{N\kappa}{T_0} \int_{-h/2}^{h/2} dy\int_{0}^{L_x} dx \left(\partial_y T^{(1)}\right)^2.
\end{align} 
The resulting first order expressions in the limit of $h/\xi\ll1$ are given in Eq.~\eqref{eq: square_scaling_th}.

The total power dissipation is subsequently:
\begin{align}
    Q_\text{tot} &= a Q_{\text{th},0}    \left(\frac{h}{\xi}\right)^3 +
     b Q_{\text{vis},0}  \left(\frac{\xi}{h}\right)^3, \\
     Q_{\text{th},0} &\equiv \frac{I^2 L_x \overline{T}}{e^2 L_y \kappa} (m \overline{\delta s})^2, \\
     Q_{\text{vis},0} &\equiv \frac{ I^2 L_x m \nu}{e^2L_y \overline{n} \xi^2},
\end{align}
where $a$ and $b$ are numerical coefficients [see Eqs.~\eqref{eq: square_scaling_th} - \eqref{eq: coeffs}] that depend on the form of the disorder potential. The width of the channel, $h$, is then determined to be that which minimizes $Q_\text{tot}$:
\begin{align}
    h &= \xi \left(\frac{b Q_{\text{vis},0}}{a Q_{\text{th},0}}\right)^{1/6} = A \xi \alpha^{-1/6} &&\\
    h&= \left(\frac{840}{\pi^4}\right)^{1/6} \xi \alpha^{-1/6} &&\:\:\:\:\:\:\,\mathrm{(square)}\\
    h&=\left(\frac{3780 }{\pi^4 \left(1-\frac{\overline{\delta n}}{\sqrt{6}\,n}\text{cos}(3\psi)\right)^2}\right)^{1/6} \xi \alpha^{-1/6} &&\mathrm{(triangular)}
\end{align}
Note that $h$ as written here for the square potential is the expression used to determine the width of the parabolic profiles in Fig.~\ref{fig: variational}. With the approximated width of the channel we can now solve for the resistivity, $\rho  = (L_y/L_x) Q_\text{tot}/I^2$, to obtain Eq.~\eqref{eq: numerical resistivity} with numerical prefactors.

\bibliography{biblio}

\end{document}